\documentclass[epjCONF, onecolumn]{svjour}
\usepackage{graphics}
\usepackage{graphicx}
\usepackage[varg]{txfonts} % Times fonts
\usepackage[latin1]{inputenc}
\session-title{Assembling the puzzle of the Milky Way}
\begin{document}
\title{Constraining the Milky Way potential using the dynamical kinematic substructures}
\author{G. Monari\thanks{\email{monari@astro.rug.nl}} 
\and T. Antoja 
\and A. Helmi }
\institute{Kapteyn Astronomical Institute, University of Groningen, 
		      P.O. Box 800, 9700 AV Groningen, The Netherlands}
\abstract{
We present a method to constrain the potential of the non-axisymmetric components of the Galaxy using the kinematics of stars in the solar neighborhood. The basic premise is that dynamical substructures in phase-space (i.e. due to the bar and/or spiral arms) are associated with families of periodic or irregular orbits, which may be easily identified in orbital frequency space. We use the ``observed" positions and velocities of stars as initial conditions for orbital integrations in a variety of gravitational potentials. We then compute their characteristic frequencies, and study the structure present in the frequency maps. We find that the distribution of dynamical substructures in velocity- and frequency-space is best preserved when the integrations are performed in the ``true'' gravitational potential.
} %end of abstract
\maketitle
\section{Introduction }
\label{sec:introduction}
The bar and the spiral arms are two examples of  non-axisymmetric features in the Galaxy. Their exact properties are not well known. Some of these may be constrained from their effect on the velocity distribution of stars near the Sun (e.g. \cite{Dehnen2000}, \cite{Fux2001}, \cite{Antoja2009}, \cite{QuillenMinchev2005}).  Here we consider toy models of the Milky Way that include a bar with pattern speed $\Omega_{\mathrm{b}} = 49\ \mathrm{km/(sec\ kpc)}$, with an angle of 20 degrees with respect to the line Sun - Galactic Center and an axisymmetric logarithmic potential with a 220  km/sec flat rotation  curve, following \cite{Dehnen2000} and \cite{Fux2001}. The toy model has been set up by generating initial conditions for $N$ particles distributed in a 2D disk embedded in this potential. These have been integrated using the Bulirsch-Stoer algorithm in a reference frame co-rotating with the bar with a conservation of Jacobi Integral $|\Delta E_J/E_J|<10^{-9}$, for sufficient time to have stationary structures in the velocity plane of the ``solar neighbourhood''.
  
\section{Characterization of the solar neighbourhood}
\label{sec:sn}
\begin{figure}
\centering
\minipage{0.45\textwidth}
  \includegraphics[width=0.9\linewidth]{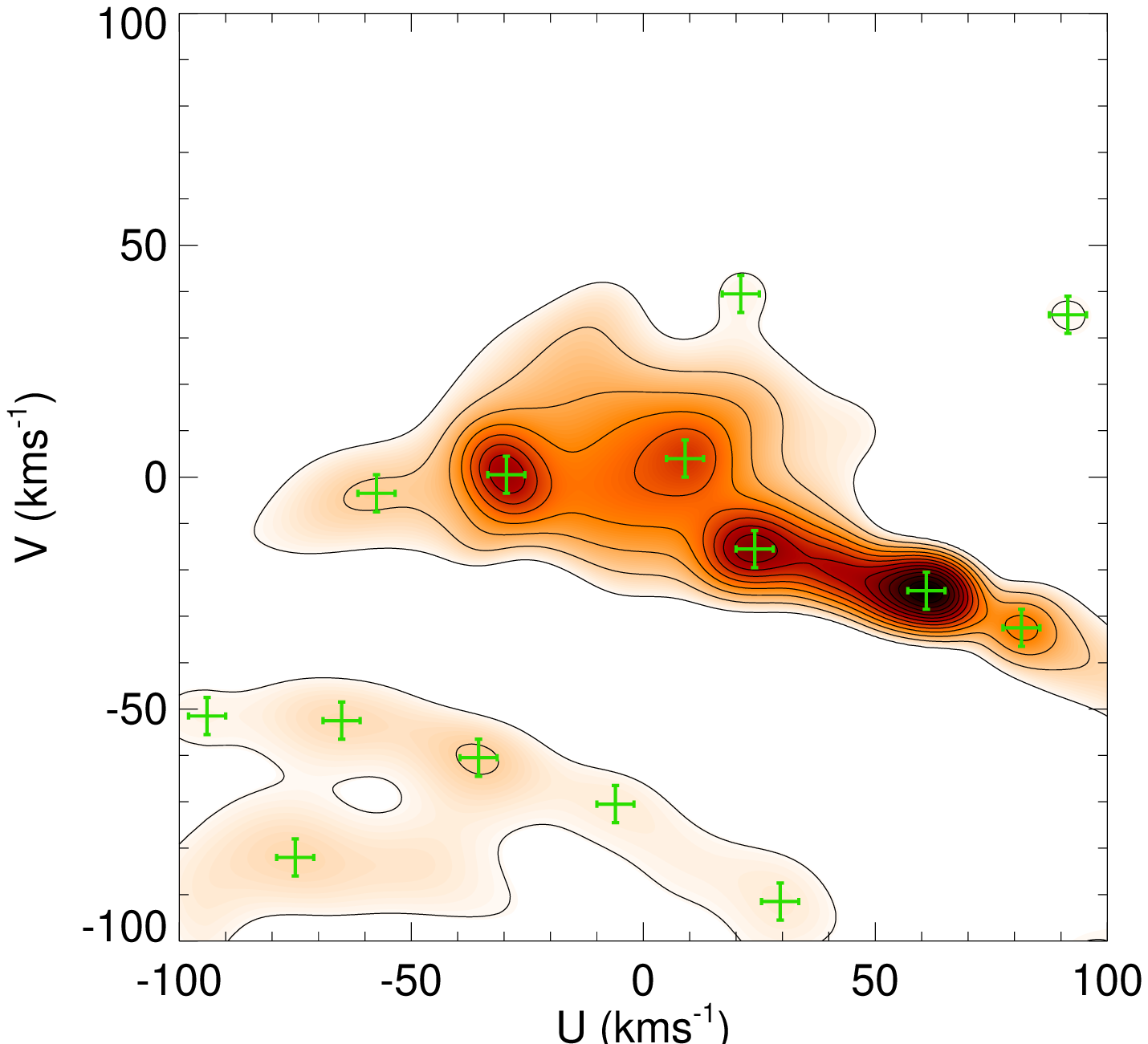}
\endminipage\hfill
\minipage{0.45\textwidth}%
  \includegraphics[width=0.7\linewidth,angle=270]{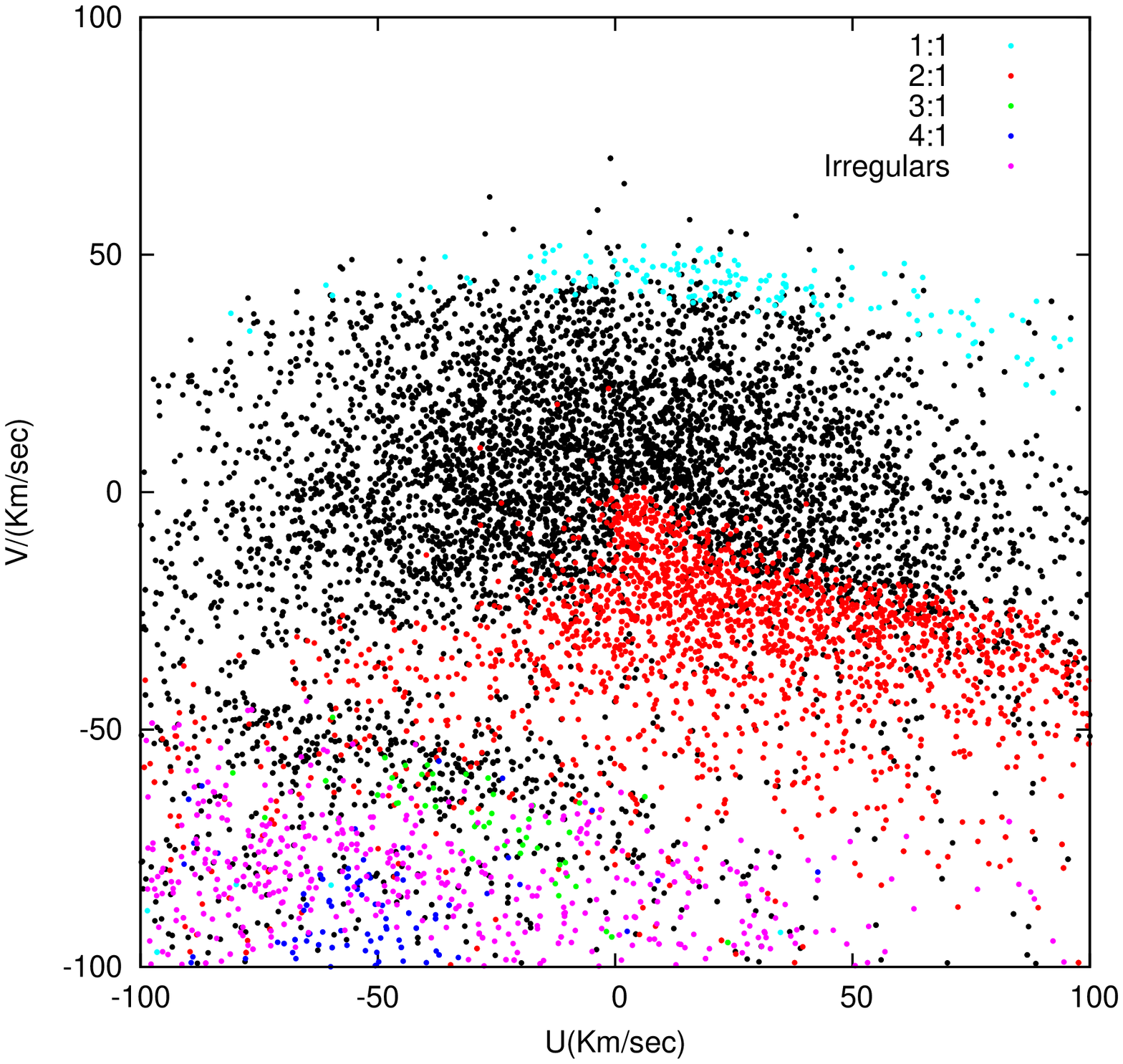} 
\endminipage
\caption{Left panel: density of ``stars'' in the velocity plane neighborhood (200 pc) obtained with a wavelet transform (described in \cite{Antoja2009}), for our toy model of a barred Milky Way.
Right panel: ``Stars'' in the velocity plane of the left panel, color-coded according to the orbit families shown in the left panel of fig. \ref{fig:2}.}
\label{fig:1}       % Give a unique label
\end{figure}

We focus on the motions and orbits of particles crossing a ``solar neighbourhood'' volume. To find the frequencies of an orbit we proceed as follows.
We compute the discrete Fourier transform of the time sequence of the radial and azimuthal coordinates of an orbit (cf. \cite{CevKly2007}). The highest peaks in the Fourier spectra correspond to the orbit's principal frequencies. When the ratio between the principal frequencies is a rational number, the orbit is said to be ``resonant''. This method is capable also to classify strongly irregular orbits.
The kinematics in the ``solar neighbourhood'' of our toy model is far from being smooth (fig. \ref{fig:1}, left panel): substructures (moving groups) can be observed. These moving groups are populated by families of resonant or irregular orbits, as shown by the frequency map (fig. \ref{fig:2}, left panel).

\section{Frequency maps to constrain the potential}
\label{sec:frqcon}
\begin{figure}
\centering
\minipage{0.45\textwidth}
  \includegraphics[width=0.7\linewidth,angle=270]{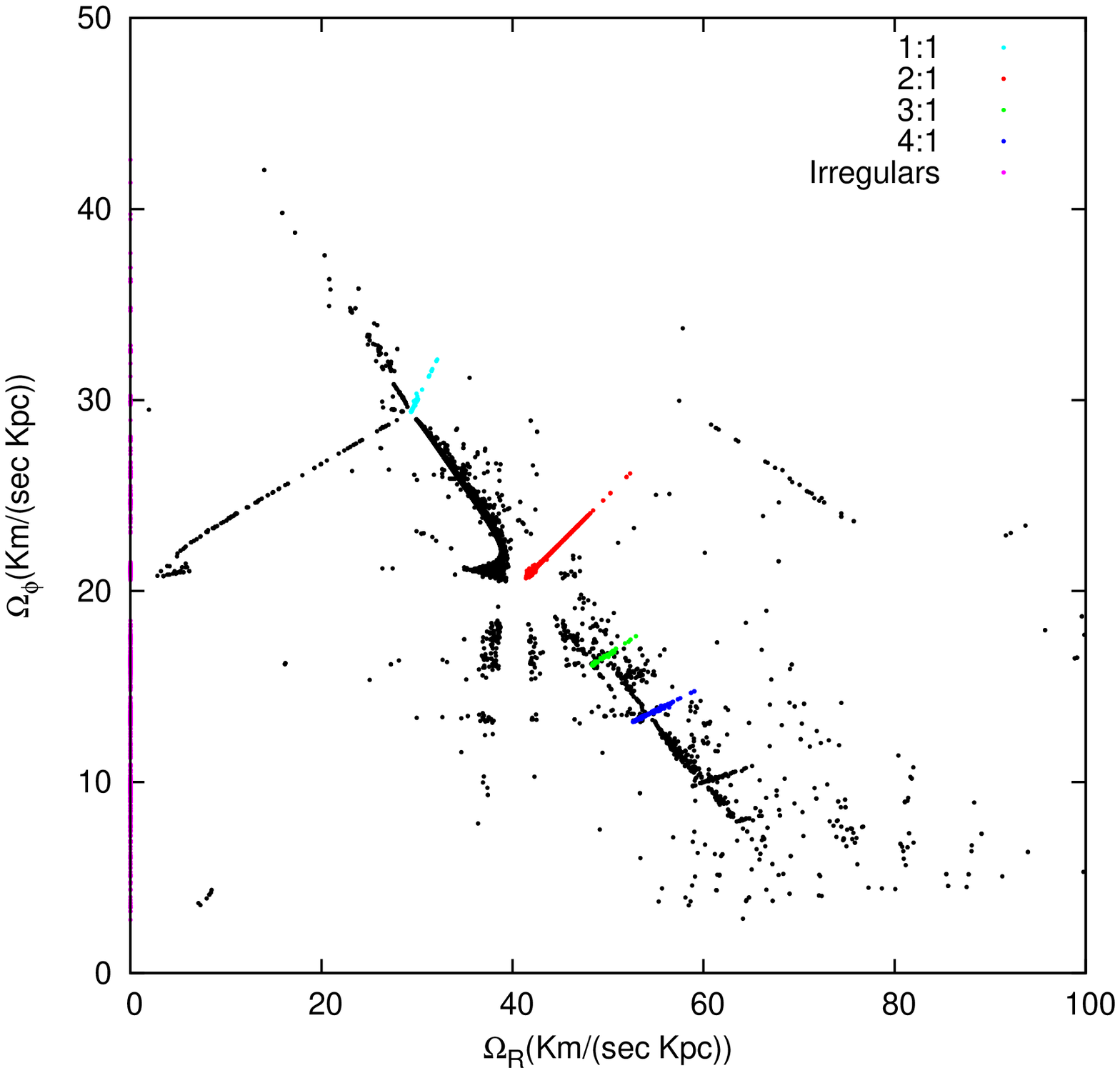}
\endminipage\hfill
\minipage{0.45\textwidth}
  \includegraphics[width=0.75\linewidth,angle=270]{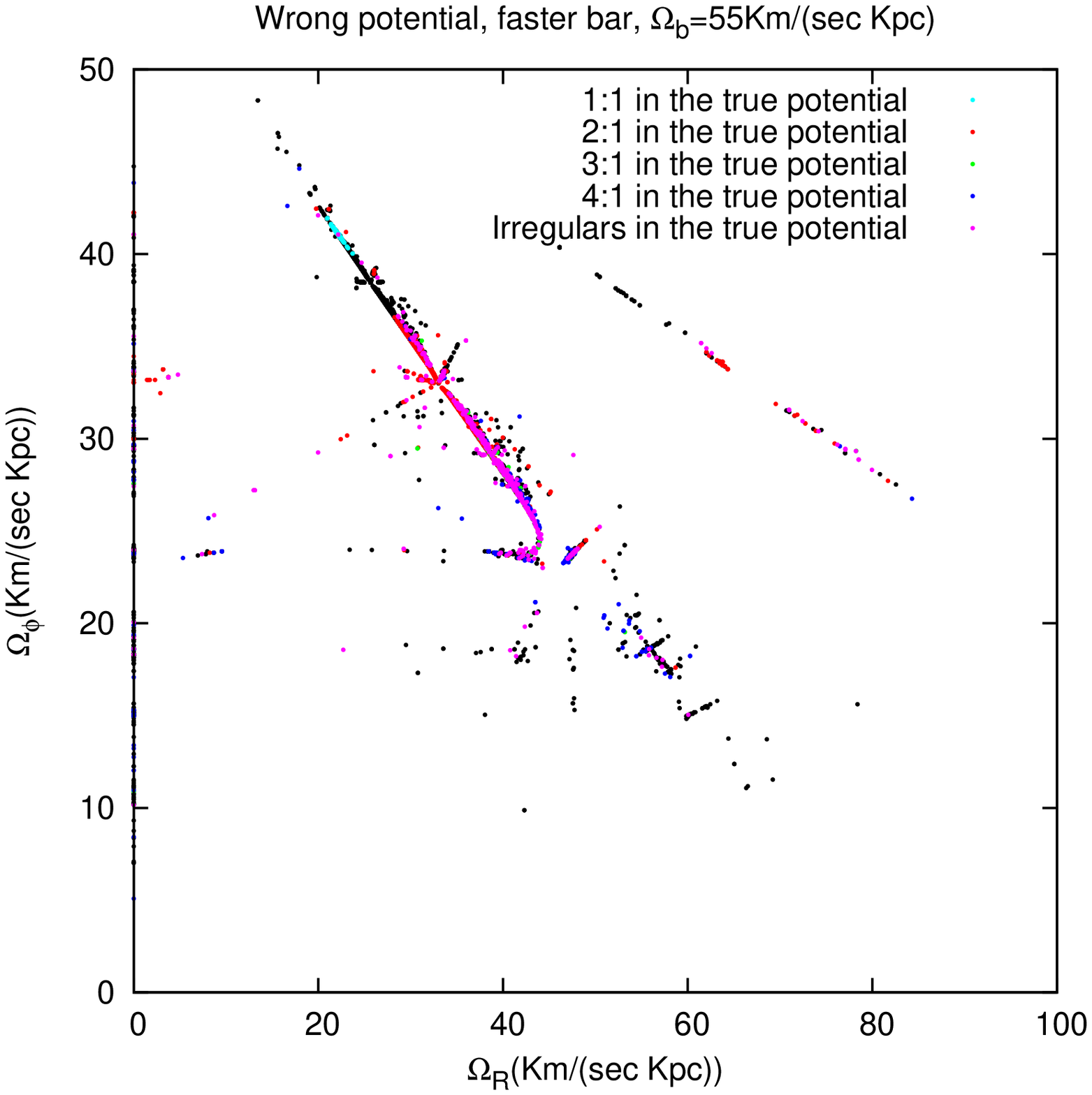}
  \endminipage\
\caption{Left: frequency plane relative to the ``stars'' of fig. \ref{fig:1}, obtained with the true potential. The different colors indicate families of irregular or resonant orbits. Note that these define straight lines in the frequency plane. Right: Example of a frequency map obtained by integrating the observed positions and velocities of  ``stars'' in fig. \ref{fig:1} in a different and wrong potential (a faster bar, in this case). The color coding corresponds to the same points of the velocity plane in the right panel of fig. \ref{fig:1}.}
\label{fig:2}       % Give a unique label
\end{figure}

We obtain frequency maps (e.g. fig. \ref{fig:2}) by integrating the observed positions and velocities of ``stars'' in fig.\ref{fig:1} in different  potentials (i.e. with varying characteristic parameters). The substructure in velocity-space appears to be  better maintained in frequency space for the right potential (fig. \ref{fig:2}, left panel). In the frequency maps with ``wrong'' potentials points originally on resonant lines are now mixed.
To quantify the degree of clustering and order in the 4D space $(U,V,\Omega_R,\Omega_{\Phi})$ we measure the information entropy for the various potentials explored. Fig. \ref{fig:2} shows that the structure of this space is quite different for different parameters, and this is also reflected in the value of the entropy.

\end{document}